\documentstyle[12pt,epsfig]{article}

\date{}

\begin{document}

\title{{\LARGE\sf Finite-Dimensional Spin Glasses:  States, Excitations,
and Interfaces}}
\author{
{\bf C. M. Newman}\thanks{Partially supported by the 
National Science Foundation under grant DMS-01-02587.}\\
{\small \tt newman\,@\,cims.nyu.edu}\\
{\small \sl Courant Institute of Mathematical Sciences}\\
{\small \sl New York University}\\
{\small \sl New York, NY 10012, USA}
\and
{\bf D. L. Stein}\thanks{Partially supported by the 
National Science Foundation under grant DMS-01-02541.}\\
{\small \tt dls\,@\,physics.arizona.edu}\\
{\small \sl Depts.\ of Physics and Mathematics}\\
{\small \sl University of Arizona}\\
{\small \sl Tucson, AZ 85721, USA}
}

\maketitle

\begin{abstract}

We discuss the underlying connections among the thermodynamic properties of
short-ranged spin glasses, their behavior in large finite volumes, and the
interfaces that separate different pure states, and also ground states and
low-lying excitations.

\end{abstract}


\small
\renewcommand{\baselinestretch}{1.25}
\normalsize

\section{Introduction}
\label{sec:intro}

In this talk, we focus on the following question: assuming that there
exists in finite dimensions a low-temperature spin glass phase with broken
spin-inversion symmetry, what are the relationships among the possible
types of ground states that might be present, their low-energy excitations,
and the interfaces that separate those states?  Answering this question
provides a handle on the broader nature of the low-temperature spin glass
phase, if it exists.

While our results apply to a wide class of short-ranged models, we focus
here for specificity on the Edwards-Anderson (EA) Ising model \cite{EA75},
with Hamiltonian:

\begin{equation}
\label{eq:EA}
{\cal H}_{\cal J}=-\sum_{<x,y>} J_{xy} \sigma_x\sigma_y -h\sum_x\sigma_x\ ,
\end{equation}
where $x$ is a site in a $d$-dimensional cubic lattice, $\sigma_x=\pm 1$ is
the Ising spin at site $x$, $h$ is an external magnetic field, and the
first sum is over nearest neighbor pairs of sites.  To keep things simple, we
take $h=0$ and the spin couplings $J_{xy}$ to be independent Gaussian
random variables whose common distribution has mean zero and variance one.
The absence of a field and the symmetry of the coupling distribution
results in an overall global spin inversion symmetry.  We denote by ${\cal
J}$ a particular realization of the couplings, corresponding physically to
a specific spin glass sample.

It is important to consider both equilibrium and nonequilibrium properties
arising from (\ref{eq:EA}), but we consider here only the former.  We
remain far from a comprehensive theory of the statistical mechanics of the
EA Hamiltonian, but nonetheless there has in recent years been substantial
analytical --- and even some rigorous --- progress in understanding what
can and cannot occur in a putative low-temperature spin glass phase.  In
this talk we examine the relationships among three properties of this
phase:

\smallskip

$\bullet$ {\bf Interfaces between ground states.}  A ground state is an
infinite-volume spin configuration whose energy cannot be lowered by the
flip of any finite subset of spins.  It can be constructed as the limit of
a sequence of finite-volume ground states, i.e., the lowest-energy spin
configuration(s) consistent with some boundary condition in a finite
volume.  The {\it interface\/} between two ground states is the set of all
couplings that are satisfied in one and not the other.

\smallskip

$\bullet$ {\bf Thermodynamic volume ($V\to\infty$) behavior.}

\smallskip

$\bullet$ {\bf Behavior in typical large finite volumes.}

\smallskip 

Our emphasis will be on the interconnectedness of these three: conclusions
about any one can provide important information about the others.  However,
these relationships may not be at all straightforward; one theme emerging
from our work is that they are far more complicated than in homogeneous
systems.

\section{`Observable' vs.~`Invisible' States}
\label{sec:obs}

How might one ``see'' the structure of the spin glass phase?  There are
several possible procedures.  As in many other statistical mechanical
problems, one can study the effects of a change in boundary conditions.
Consider a cube $\Lambda_L$ of side $L$ centered at the origin with
periodic boundary conditions.  There will a spin-flip-related pair of spin
configurations within $\Lambda_L$ of lowest energy.  This finite-volume
ground state pair will change in general with the boundary condition; for
example, if one switches to antiperiodic boundary conditions, then there
will be a new ground state pair with some interface relative to the old
one.

One important question is whether such interfaces are {\it pinned\/} or
{\it deflect to infinity\/}: if for any fixed $L_0$, the relative interface
eventually moves (and stays) outside of $\Lambda_{L_0}$ as $L\to\infty$,
the interface has `deflected to infinity'; if it remains inside
$\Lambda_{L_0}$, it is `pinned' (for an illustration, see Fig.~2 of
\cite{NS02}).  Pinned interfaces imply the existence of infinite-volume
multiple ground state pairs.

More recent techniques are useful for studying excitations whose energies
above a finite-volume ground state are of $O(1)$ independent of the volume
size.  Consider again a cube $\Lambda_L$ with periodic boundary conditions.
The method of Krzakala and Martin (KM) \cite{KM00} is to choose an arbitrary
pair of spins, and force them to have an orientation opposite to that in
the ground state pair.  The method of Palassini and Young (PY) \cite{PY00} is
to add a perturbation that lowers the energy of a spin configuration by an
amount proportional to the fraction of bonds in an interface relative to
the ground state pair.  Either way, there will be a new spin configuration
pair that minimizes the energy.

What these and most other widely-used procedures have in common is that
they are not explicitly coupling-dependent; e.g., the boundary conditions
used are independent of the coupling realization ${\cal J}$.  We call
`observable' any interfaces (and resulting states) that can be constructed
in this way; interfaces or states that can only be seen using coupling-{\it
dependent\/} boundary conditions we call `invisible'.  A detailed
discussion of the reasons behind these designations is given in
\cite{NSinprep} (see also the discussion in Sec.~3 of~\cite{vEF85}).

\section{Interfaces}
\label{sec:interfaces}

We now focus on interfaces separating spin configurations.  The analysis is
at $T=0$ but can be extended to nonzero temperature.  That is, we confine
our attention here to interfaces separating either different ground states
or ground states and excitations.  For now, we won't worry how the
interfaces arise: they might have arisen between ground states in a single
volume under a switch from periodic to antiperiodic boundary conditions, or
from using the KM or PY procedures, or through some other
coupling-independent method.  Our discussion below applies to all.

The main features we study include:

\smallskip

$\bullet$ {\bf Spatial Structure.}  In other words, is the interface
`space-filling' or `zero-density'?  By space-filling we mean the following.
Consider in $d$ dimensions a sequence of cubes $\Lambda_L$, each containing
an interface generated through a common procedure (e.g., switching from
periodic to antiperiodic boundary conditions).  If, for each $L$, the
number of bonds in the interface scales as $L^{d_s}$ with $d_s=d$, then the
interface is space-filling.  If $d_s<d$, it is zero-density.

One of the interesting possibilities arising in spin glasses is the
possibility of space-filling interfaces separating ground or pure states;
such a possibility cannot arise, e.g., in ferromagnets.  Space-filling
interfaces (suitably redefined) are believed \cite{MPV87} to separate
different pure states in the low-temperature phase of the infinite-ranged
Sherrington-Kirkpatrick (SK) model \cite{SK75}.

\smallskip

$\bullet$ {\bf Energetics.}  Spin glass interfaces can also have unusual
energetic properties.  In a ferromagnet, whether homogeneous or disordered,
the energy of an interface scales linearly with the number of bonds
comprising it.  This is not necessarily so in a spin glass.  One intriguing
possibility is that --- as happens in the SK model --- an interface might
have $O(1)$ energy independently of its size.  The other possibility is
that the energy of an interface in the volume $\Lambda_L$ is $O(L^\theta)$,
with $\theta>0$.  (We ignore unlikely special cases such as logaritihmic
and other non-power-law dependences.)

\smallskip

$\bullet$ {\bf Pinning.}  A property crucial to the nature of states
separated by an interface is whether the interface is pinned or deflects to
infinity, as discussed in Sec.~\ref{sec:obs}.  As shown in
\cite{NS02,NS01}, which of these occurs is not independent of the spatial
structure, for interfaces separating observable states: a space-filling
interface must be pinned, and a zero-density one (generated by a
coupling-independent procedure) must deflect to infinity.

\smallskip

\begin{figure}[t]
\centerline{\epsfig{file=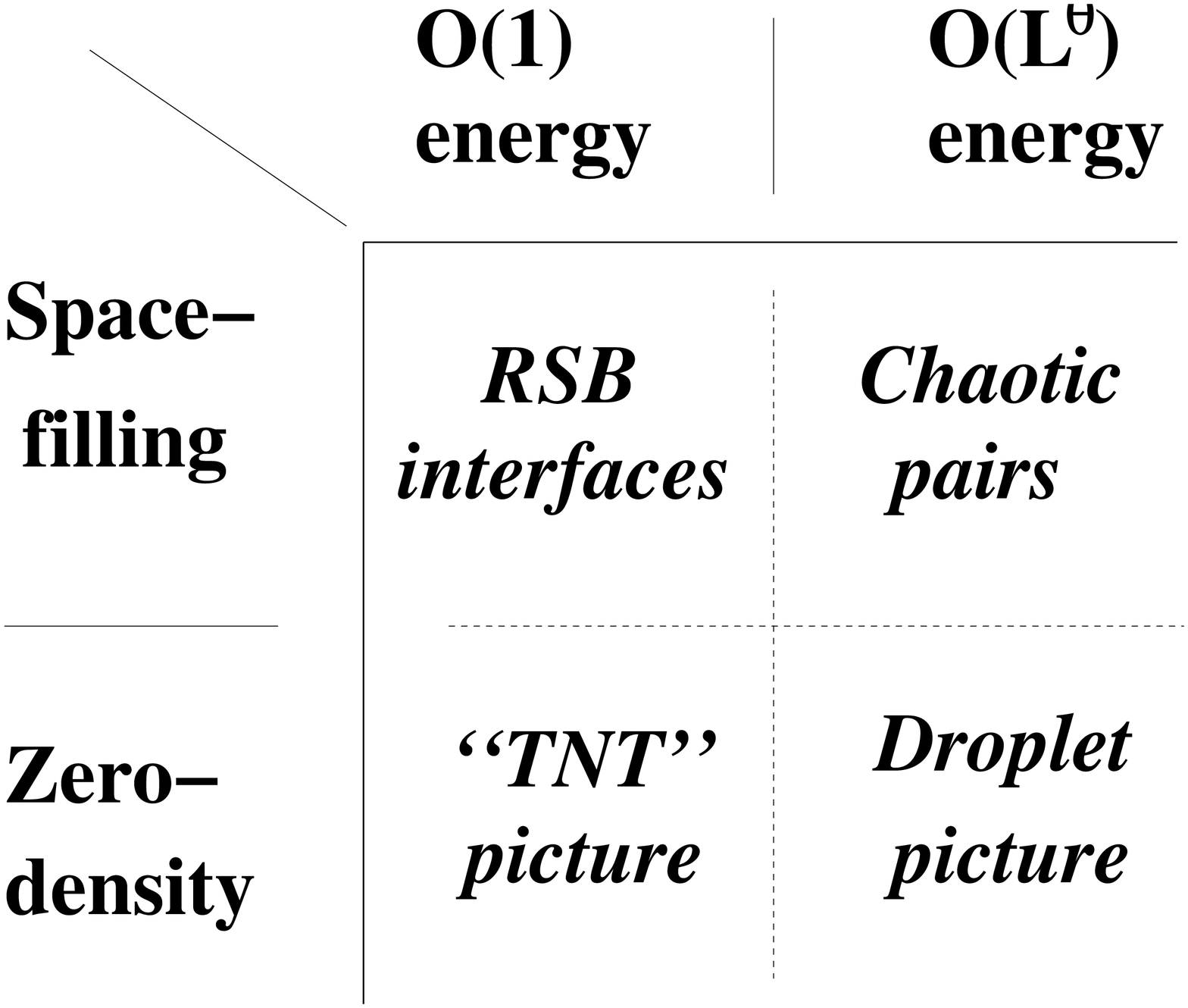,width=4.0in}}
\renewcommand{\baselinestretch}{1.0} 
\caption{Table illustrating the correspondence between a type of interface
and a scenario for the structure of the low-temperature spin glass phase
in finite dimensions.}
\label{fig:table}
\end{figure}

Each of these interface properties corresponds to a major scenario proposed
for the low-temperature spin glass phase of the EA model in finite
dimensions.  This is summarized in Fig.~\ref{fig:table}, and each will be
discussed in turn.

\renewcommand{\baselinestretch}{1.25}
\normalsize

\subsection{Mean-Field Picture}
\label{subsec:mfp}

The picture arising from Parisi's solution
\cite{MPV87,P79,P83,MPSTV84a,MPSTV84b} of the infinite-ranged SK model
is known as the replica symmetry breaking (RSB) theory.  There have
been many papers written in support of the notion that the RSB theory
should apply as well to short-ranged spin glasses.  We will not
discuss all of RSB's features here; given the subject of this talk, we
focus on the prediction \cite{MP00a,MP00b} of RSB theory that
interfaces separating ground states are both space-filling and can
have energies of $O(1)$, i.e., that don't scale with the size of the
system.  Hence the designation `RSB interfaces' in the upper left-hand
corner of the table in Fig.~\ref{fig:table}.

A central result of \cite{NS02} was a proof that the existence of
space-filling interfaces (regardless of how their energies scale, as long
as the interfaces are generated by coupling-independent methods) is a
sufficient condition for the existence of multiple thermodynamic pure state
pairs.  In order to have a nontrivial overlap function (at nonzero
temperature) of the sort characteristic of RSB theory \cite{MPV87}, we have
shown that the set of all states generated through sequences of
finite-volume Gibbs states must partition into a union of thermodynamic
{\it mixed\/} states $\Gamma$, with each $\Gamma$ being a nontrivial
collection of infinitely many pure states with certain weights.  Different
$\Gamma$'s would appear in different finite volumes; for details, see
\cite{NS96b,NSBerlin,NS97,NS98}.  One can also prove that there must be an
{\it uncountable\/} infinity of pure states in the union of all $\Gamma$'s
\cite{NSunpub}.

So in this picture the connections among the three properties listed in
Sec.~\ref{sec:intro} is that RSB interfaces imply (and are implied by) this
complex thermodynamic structure, and by nontrivial overlap structure in
large finite volumes.  However, thermodynamic arguments show
\cite{NS02,NS96b,NSBerlin,NS97,NS98,NS96a} that this thermodynamic
structure cannot be supported in finite-dimensional, short-ranged spin
glasses, in turn ruling out the possibility that space-filling, $O(1)$
energy interfaces can arise in these systems.

\subsection{Chaotic Pairs Picture}
\label{subsec:cp}

The upper right-hand corner of Fig.~\ref{fig:table} raises the possibility
of space-filling interaces whose energy scales as $L^\theta$, with
$\theta>0$.  (For reasons that won't be discussed here, there is an upper
bound \cite{FH86,NS92} of $\theta\le (d-1)/2$ for observable states.)  This
leads directly to the {\it chaotic pairs\/} picture
\cite{NS96b,NSBerlin,NS97,NS98,NS92} in which many pure state pairs exist,
but only a single spin-reversed pair appears in any large finite volume,
leading to a trivial overlap structure and no ultrametricity or related
features of the RSB scenario.

\subsection{Droplet/Scaling Picture}
\label{subsec:ds}

Continuing clockwise around the table in Fig.~\ref{fig:table}, we come to
interfaces that are zero-density and with energy scaling as $L^\theta$,
with $\theta>0$.  These interfaces characterize droplet excitations in a
two-state picture developed by Macmillan \cite{Mac84}, Bray and Moore
\cite{BM87}, and Fisher and Huse \cite{FH86}.  This picture is well-known
and won't be discussed in detail here.

\subsection{`TNT' Picture}
\label{subsec:TNT}

We come finally to the last entry, which conjectures interfaces that are
both zero-density and whose energies remain $O(1)$ independently of
lengthscale.  This picture was proposed by Krzakala and Martin \cite{KM00}
(who denoted the picture `TNT' for trivial link overlap and nontrivial spin
overlap) and Palassini and Young \cite{PY00}.  It can be shown that
zero-density interfaces cannot separate observable pure or ground states
\cite{NS01}, so these states, should they exist, would necessarily be
excitations, as in the droplet-scaling picture.

\section{Conclusion}
\label{sec:conclusion}

The purpose of this talk has been to demonstrate that there are deep
connections among the thermodynamic structure of spin glass states, their
behavior in large finite volumes, and the interfaces that separate these
states (and/or their excitations).  Tools developed for any one can lead to
strong conclusions about the other two.  This interconnection has allowed
substantial analytical, and even rigorous, progress on short-ranged spin
glasses.

\medskip

{\it Acknowledgments.}  We thank the organizers of this conference, Daniel
Iagolnitzer and Jean Zinn-Justin, for their invitation to speak and for
their hard work in putting together a very enjoyable meeting.

\renewcommand{\baselinestretch}{1.0}

\end{document}